\documentclass[aps,prd,reprint,showpacs]{revtex4-1}
\pdfoutput=1
\usepackage{color}
\usepackage{amsmath}
\usepackage{amssymb}
\usepackage{amsfonts}

\newcommand{\doctitle}{Revisiting the implications of CPT and unitarity for baryogenesis and leptogenesis}
\newcommand{\auth}{Atri Bhattacharya, Raj Gandhi, Satyanarayan Mukhopadhyay}

\usepackage[
            pdftitle={\doctitle},
            pdfkeywords={CPT, Leptogenesis, Baryogenesis},            
            pdfauthor={\auth},   
           ]{hyperref}      
\usepackage[all]{hypcap}
\usepackage{graphicx}
\usepackage{array}
\usepackage{slashed} 
\usepackage{hyphenat}
\usepackage[normalem]{ulem}
\usepackage{dcolumn} 
\usepackage{paralist}
\usepackage{mathrsfs} 
\usepackage[shadow,textsize=footnotesize]{todonotes}

\newcommand{\wn}{Nanopoulos\hyp{}Weinberg }
\newcommand{\ket}[1]{\mid #1 \rangle}
\newcommand{\bra}[1]{\langle #1 \mid}
\newcommand{\abs}[1]{\big\lvert #1 \big\rvert}
\newcommand{\bigabs}[1]{\bigg\lvert #1 \bigg\rvert}
\newcommand{\ie}{\textit{i.e.}}
\newcommand{\eg}{\textit{e.g.}}

\newcommand{\anti}[1]{\ensuremath{\bar{#1}}}
\newcommand{\Lag}{\ensuremath{\mathscr{L}}}
\newcommand{\slb}{\ensuremath{\slashed{B}}}
\newcommand{\aslb}{\ensuremath{\alpha_\slb}}
\newcommand{\tilt}{\ensuremath{\widetilde{T}}}
\newcommand{\tilta}{\ensuremath{\tilt_1}}
\newcommand{\tiltb}{\ensuremath{\tilt_2}}
\newcommand{\sumb}[2]{\ensuremath{\sum_{#1\in #2}}}
\newcommand{\bigo}{\ensuremath{\mathcal{O}}}

\begin{document}

\title{\doctitle}

\author{Atri Bhattacharya}
\email{atrib@email.arizona.edu}
\affiliation{Department of Physics, University of Arizona, Tucson, AZ 85721, USA}
\author{Raj Gandhi}
\email{nubarnu@gmail.com}
\affiliation{Harish-Chandra Research Institute, Chhatnag Road, Jhunsi, Allahabad-211019, India}
\author{Satyanarayan Mukhopadhyay}
\email{satya.mukho@ipmu.jp}
\affiliation{Kavli IPMU (WPI), The University of Tokyo, Kashiwa, Chiba 277-8583, Japan}


\begin{abstract}
In the context of GUT baryogenesis models, a well-known theorem asserts that CPT conservation and the unitarity of S-matrix require that the lowest order contribution that leads to the generation of  a non-zero  net CP-violation via the decay of a heavy particle must be to $\bigo(\aslb^3)$, where $\aslb$ is a baryon number (B) violating coupling.  We revisit this theorem (which holds for lepton number (L) violation, and hence for leptogenesis as well) and examine its implications for models where the particle content allows the heavy particle to also decay via modes which conserve B (or L) in addition to modes which do not.  We  systematically expand the S-matrix order by order in B\slash L-violating couplings, and show, in such cases, that the net CP-violation is  non-zero even  to $\bigo(\aslb^2)$, without actually contradicting the theorem. By replacing a B/L violating coupling (usually constrained to be small) by a relatively unconstrained B/L conserving one,  our result may allow for sufficient CP  violation in  models where  it may otherwise have been difficult to generate the  observed baryon asymmetry.  As an explicit application of this result, we construct a model in low-scale leptogenesis.
\end{abstract}

\maketitle

\section{\label{sec:intro}Introduction}
The asymmetry in the universe between baryonic and anti-baryonic matter is expressed in terms of the ratio,
\begin{equation}
	Y_{\Delta B} \equiv \frac{n_B - n_{\anti{B}}}{s},
\end{equation}
where, $n_B$ and $n_{\anti{B}}$ represent the baryon and anti-baryon densities respectively, and $s=g_*(2\pi^2/45)T^3$ is the entropy density, $g_*$ the number of relativistic degrees of freedom in the plasma, and $T$ is the temperature. 
The current estimate for this asymmetry has been determined independently from i) the abundances of light nuclei due to big bang nucleosynthesis(BBN) and, ii) analyses of the Cosmic Microwave Background Radiation (CMB). Its  values (at 95\% C.L.)\cite{Iocco:2008va,Larson:2010gs},
\begin{equation}
\begin{split}
Y_\text{BBN} &= (8.10\pm0.85)\times 10^{-11}\,,\\
Y_{CMB} &= (8.79\pm0.44)\times 10^{-11}
\end{split}
\end{equation}
 confirm that we exist in a universe that is baryon dominated \cite{Steigman:1976ev,Cohen:1997ac}. The consistency beween these independent measurements of the baryon asymmetry is all the more impressive because their respective epochs are separated by about six orders of magnitude in temperature, putting its existence on a firm experimental footing.

At variance with this, however, is the fact that  a largely symmetric universe, in terms of matter and anti-matter, is expected from our present theoretical understanding of the early universe and the extremely tiny amount of matter-antimatter asymmetry present  in the quark sector of  fundamental particle interactions. While B violation, the first of the well-known Sakharov conditions~\cite{Sak} for the generation of the asymmetry may well be realized at high temperatures in the early universe~\cite{Kuzmin:1985mm}, the second condition of CP violation~\footnote{C, P and T will denote the charge-conjugation, parity and time-reversal transformations, respectively, henceforth in our work.} requires a mechanism beyond the Kobayashi-Maskawa complex phase \cite{Kobayashi:1973fv} of the Standard Model. Similarly, the third Sakharov condition of departure from thermal equilibrium may require extending the physics of the Standard Model. The latter   allows non-equilibrium processes to occur at the electro-weak phase transition \cite{Huet:1994jb,Rubakov:1996vz}, but these may not be sufficiently first-order  and thus unable to generate the requisite asymmetry \cite{Kajantie:1995kf}. It is thus fair 
to say that while several interesting theories have been proposed to explain the dynamical generation of this asymmetry, the actual mechanism by which this occurs in nature remains to be established. 

Baryogenesis is a class of  mechanisms that attempt to explain the asymmetry  by postulating its dynamic generation in the early universe, during the period between the end of cosmological inflation and reheating, and prior to the onset of nucleosynthesis, via interactions of particles and anti-particles asymmetric in their rates (see refs.~\cite{Cline:2006ts,Riotto:1999yt,Riotto:1998bt} for detailed reviews). Examples of mechanisms which have been proposed include a) GUT Baryogenesis models \cite{Ignatiev:1978uf,Yoshimura:1978ex,Toussaint:1978br,Dimopoulos:1978kv,Ellis:1978xg,
Weinberg:1979bt,Yoshimura:1979gy,Barr:1979ye,WN,Yildiz:1979gx}, b) Electroweak baryogenesis  \cite{,Rubakov:1996vz,Riotto:1999yt,Cline:2006ts} , c) the Affleck-Dine mechanism \cite{Affleck:1984fy} and d) Spontaeneous baryogenesis \cite{Cohen:1987vi,Cohen:1988kt}.
 In recent times, however, much attention has been focussed on achieveing Baryogenesis via Leptogenesis \cite{FY}. This involves the initial generation of an asymmetry in the lepton-antilepton content of the universe and its subsequent conversion to baryon asymmetry by means of sphaleron interactions that violate baryon (\emph{B}) and lepton (\emph{L}) numbers simultaneously, while conserving $B-L$ (see refs.~\cite{Fong:2013wr,Nardi:2013mha,Pilaftsis:2013nqa,Pilaftsis:2009pk,DiBari:2012fz,DNN,BPY} for  reviews on the subject). 

Our  work focuses on the constraints that are imposed on models of baryogenesis (including baryogenesis via leptogenesis)  by the  fundamental invariances of CPT and unitarity in quantum field theories.

\paragraph*{}
The general consequences of CPT-invariance and unitarity of the S-matrix in the context of the generation of baryon asymmetry in GUT models have been explored in the past~\cite{WN,KW}. 
In particular, as first pointed out by Nanopoulos and Weinberg~\cite{WN}, while calculating the CP-asymmetry generated in B-violating heavy particle decays, the leading contribution to the asymmetry involves processes which are to the third-order or higher in the B-violating coupling$\aslb$. 
Thus the (amplitude-level) contribution of graphs to the first order in B\slash L (\ie~B or L) violation (and to all orders in B/L conserving interactions) vanishes as a consequence of CPT invariance and unitarity of the S-matrix in the theory. Henceforth, we shall refer to this result as the \wn theorem. Its importance lies in it being a  general result that applies to any particle physics model that attempts to dynamically generate the baryon asymmetry and the requisite CP violation with interaction vertices  that break B or L. The most  significant application has been to non-equilibrium decays of heavy particles which constitute the spectrum of theories beyond the standard model. 

Although widely applicable, the \wn theorem was formulated in the context of  massive guage boson decays associated with  GUTs. An important input in proving this theorem was that, in the models considered in ref.~\cite{WN}, all decay modes of these heavy bosons were B-violating. 
Such an assumption is, of course, completely justified when formulating a minimal model satisfying the requirements for GUT-based baryogenesis. However, we note that in the present context of efforts to carry physics beyond the standard model, a wide range of possible models with varying particle content which can provide the seeds for B\slash L generation have been studied in the literature. In this wider framework, the heavy particle which leads to a CP asymmetry by its decay may have access to decay modes which conserve B (or L) in addition to those which violate it. Our work pertains to such cases, and points out a facet of the theorem which may guide the building of  baryogenesis and leptogenesis models which have not received adequate attention so far.

In what follows,  we  re-visit the impact of CPT and unitarity on asymmetry generating  interactions by looking at the S-matrix order-by-order in B\slash L violating couplings, and determine the leading order in these couplings at which the net CP-violation generated is non-zero. 
Specifically, we study the generic  scenarios where the parent particle has access to  a) only B\slash L violating decay modes and b) to  both B\slash L conserving and violating ones. 
The essential  upshot of our considerations is that in models where a consistent and natural scheme of B\slash L number assignment leads to the presence of both B\slash L violating and conserving decay modes of a heavy particle, the net CP-violation to $\bigo(\aslb^2)$ ( calculated with graphs to only first order in B\slash L violation) is non-zero. We  emphasize  that our result is in no way contradictory to the \wn theorem, but rather a useful re-analysis and extension,  which might be helpful while considering the building of  various models to achieve  baryogenesis and leptogenesis.

This paper is organised as follows: in section \ref{sec:cp_violation}, we review the constraints imposed by CPT invariance and unitarity of the S-matrix on the possible generation of CP violation in the decays of heavy particles. In section \ref{sec:order-exp}, we find expressions for the B\slash L asymmetries generated in different schemes of B\slash L assignment for the decaying particle and demonstrate their equivalence. We also explore the consequences of the re-formulation of the \wn theorem by constructing an example model of leptogenesis in the same section. The last section contains our conclusions.

\section{\label{sec:cp_violation}CP-violation in heavy particle decay}

\subsection{\label{ssec:cpt_unitarity}General implications of CPT invariance and S-matrix unitarity}

We first briefly review the general implications of CPT conservation and unitarity of the S-matrix for various interactions~\cite{WN,KW}.

Let us assume that the initial state of a system represented by $i$ (which represents all the quantum numbers of the system at this state) proceeds via interactions to a final state $f$. The probability of a  transition to a state $f$ from the state $i$ is given by $\abs{S_{fi}}^2$, where
\begin{equation}
	S_{fi} = \bra{f} S \ket{i}\;
\end{equation}
is the so-called S-matrix element.
This S-matrix can be decomposed as follows:-
\begin{equation}\label{eqn:S-matrix-decompose}
	S_{fi} = \delta_{fi} + \imath  T_{fi}\;,
\end{equation}
where, $T_{fi}$ represents the $fi$-th element of the \emph{T-matrix}, which represents the probability amplitude of transition of a system in the initial state $i$ to a distinct final state $f$, \ie, without transitioning to itself. 
The S-matrix must be unitary,
\begin{equation}
	S S^\dag = 1 = S^\dag S \label{eqn:sdags}.
\end{equation}
Written out in terms of the elements after inserting a complete set of states wherever necessary, this gives,
\begin{subequations}
\begin{align}
	\sum_f \abs{S_{fi}}^2 &= 1,\text{ and,}\label{eqn:Sfi_unitarity} \\
	\sum_f \abs{S_{if}}^2 &= 1.\label{eqn:Sif_unitarity}
\end{align}
\end{subequations}
Equivalently, in terms of the T-matrix this can be expressed as,
\begin{subequations}\label{eqn:unitary_T}
	\begin{align}
		\sum_n T^*_{ni} T_{nf} 
		&= -\imath \left( T_{if} - T_{fi}^* \right),			
		\label{eqn:unitary_T-0}
		\\ 
		\intertext{which implies, for $i = f$,} 
		\sum_n \abs{T_{ni}}^2 &= 2\Im \left( T_{ii} \right),
		\label{eqn:unitary_T-1}
	\end{align}	
\end{subequations}
where, in going from Eq.~\eqref{eqn:unitary_T-0} to \eqref{eqn:unitary_T-1} we have denoted the \emph{imaginary part} of the complex quantity $T_{if}$ by $\Im(T_{if})$. It is easy to show, along the same lines but starting from Eq.~\eqref{eqn:Sif_unitarity} instead of \eqref{eqn:Sfi_unitarity}, that, also
\begin{equation}\label{eqn:unitary_T2}
	\sum_n \abs{T_{in}}^2 = 2\Im \left( T_{ii} \right).
\end{equation}

Further, conservation of CPT ensures that the probability of transition of an initial state $i$ to a final state $f$ is equivalent to that of the transition of the corresponding CP conjugate states $\anti{f}$ to $\anti{i}$
\begin{equation}\label{eqn:t-matrix-cpt}
	T_{fi} = T_{\anti{i}\anti{f}}.
\end{equation}
The consequence of unitarity as expressed in Eq.~\eqref{eqn:unitary_T} and \eqref{eqn:unitary_T2}, along with  CPT invariance ensures that
\begin{equation}
		\sum_{\anti{f}} \abs{T_{\anti{f}\anti{i}}}^2 = \sum_f \abs{T_{fi}}^2. 
\end{equation}
Therefore, the probability of a system in a state $i$ transitioning to all possible final states $f$ is identical to the probability of the system in the CP conjugate state \anti{i} transitioning to all possible final states $\anti{f}$. This is an important consequence of CPT conservation and unitarity and it tells us, among other things, that the \textit{total} decay width of a particle and its CP conjugate (anti-particle) are necessarily identical.

\paragraph*{CP violating amplitudes and unitarity.}
As opposed  to constraints on  sums over all final states, as considered above, we now pose the question: what constraint does unitarity impose on individual CP-violating amplitudes? If the particular  interaction that generates the transition amplitude $T_{fi}$ is CP non-conserving, then the difference between the probabilities of the CP conjugate processes $ i \rightarrow f $ and $ \anti{i} \rightarrow \anti{f} $, or equivalently between $ i \rightarrow f $ and $ f \rightarrow i $ is finite and non-zero. Indeed, using Eq.~\eqref{eqn:unitary_T-1} in the form
\begin{equation}
	T_{if} = \imath \sum_n T_{in}T^*_{fn} + T^*_{fi},
\end{equation}
it is straightforward to obtain an expression for the difference in the probabilities for the CP conjugate interactions
\begin{equation}
\begin{split}\label{eqn:KWeqn}
	\abs{T_{\anti{f}\anti{i}}}^2 - \abs{T_{fi}}^2
	&=
	\abs{T_{if}}^2 - \abs{T_{fi}}^2 
	\\
	&= -2 \Im \biggl( \sum_n \left(T_{in} T^*_{fn}\right)T_{fi} 
              \biggr) 
    \\
	&\quad
	+ \bigabs{\sum_n T_{in}T^*_{fn}}^2
\end{split}
\end{equation}
This equation implies that CP-violating differences are generated by the interference of tree and loop graphs, where the intermediate states in the loop are on-shell \cite{KW} --- leading to a non-zero imaginary part in the amplitude.

At this juncture, it is appropriate to recall  the result of the \wn theorem, which examined the net baryon excess $\Delta B$ produced in the decays of super-heavy $X$ bosons and their anti-particles. The conclusion derived there \cite{WN} was  that graphs to first order in B-violating interactions but to arbitrary order in baryon-conserving interactions make no contribution to a net  $\Delta B$. In particular, it was shown that when decay amplitudes are calculated using graphs to first-order in B-violating interactions, CPT invariance requires that the decay rate for a particle $X$ into all final states with a given baryon number $B$  equals the rate for the corresponding decay of the anti-particle $\anti{X}$ into all states with baryon number $-B$. Therefore, this theorem indicates that one must consider graphs to at least second order in B-violating interactions.

We note, however, that  in this paper the authors considered models where the super-heavy boson giving non-zero contribution to the net baryon asymmetry had only B-violating decay modes. This assumption was incorporated in the proof of this theorem by demanding that in the absence of B-violating interactions, the wave-function of $X$, $\psi_X$ is a one-particle state. As noted in the Introduction, over the past two decades, many classes of models for baryogenesis (and leptognesis) have appeared in the literature, with particle spectra involving not just heavy GUT scale  guage bosons, but also BSM (i.e. beyond standard model)  scalars and Majorana fermions, with B/L  and CP violating interactions.  It is thus reasonable and relevant  to relax this particular assumption in the wider context of BSM models  and their particle content. By introducing decay modes which are not always B/L violating, it is expected that the result in \cite{WN} will be modified when subjected to the same constraints of CPT and unitarity. It is the study of this modification and its consequences for present day B/L violating models  which is the main objective of this paper.

In section~\ref{sec:order-exp}, to begin with we shall implement this assumption at the S-matrix level by demanding that in the case where the heavy boson $X$ decays only via B-violating interactions, the S-matrix elements  $(S_0)_{fX}=\delta_{fX}$, where $S_0$ denotes the part of the S-matrix which contains only B-conserving interactions. Expanding the S-matrix order-by-order in B-violating couplings $\aslb$, we then show that the net CP-violation generated is zero to $\bigo(\aslb^2)$, which of course, is tantamount to re-deriving the result in \cite{WN} (Case 1 in Section III below). Next, we relax the assumption and examine the  consequences (Case 2, Section III).

\section{\label{sec:order-exp}Systematic expansion of the S-matrix in B\slash L-violating couplings}

We first split the S-matrix into two parts, 
\begin{equation}\label{eqn:s-decompose-tilt}
	S = S_0 + \imath \tilt,
\end{equation}
where $S_0$ includes the identity element of the total S-matrix and also processes represented by graphs with only B-conserving interactions. \tilt~contains processes described by graphs with B-violating interactions to first order or higher and B-conserving interactions to all orders. 
Using this expansion in Eq.~\eqref{eqn:sdags} we arrive at the following relation between $S_0$ and $\tilt$
\begin{equation}
	\tilt = S_0 \tilt^\dag S_0 + \imath S_0 \tilt^\dag \tilt 
	= S_0 \tilt^\dag S
\end{equation}
In terms of the elements of the S and T matrices, we therefore see that
\begin{equation}
	\tilt_{Xf} = \sum_{i,j}\left( S_0 \right)_{Xi} \bigl( \tilt^\dag \bigr)_{ij} S_{jf}\label{eqn:main-eqn}
\end{equation}
From Eq.~\eqref{eqn:main-eqn} we get
\begin{equation}
	\abs{\tilt_{Xf}}^2 = \sum_{i,j,k,m}\left( S_0 \right)_{Xi}  \tilt_{ji}^* S_{jf} \left( S_0 \right)^*_{Xk} \tilt_{mk} S^*_{mf} \label{eqn:main-eqn2}
\end{equation}

Denoting all B-violating coupling constants by \aslb, we expand the quantity \tilt~in a perturbation series in this coupling constant
\begin{equation}
	\label{eqn:tilt-exp}
	\tilt = \aslb \tilta + \aslb^2 \tiltb + \cdots,
\end{equation}
where the quantities \tilta~and \tiltb~themselves do not contain any factors of the B-violating coupling constant \aslb.
Thus,
\begin{subequations}\label{eqn:s0-exp}
\begin{equation}
	S = S_0 + \imath\left(\aslb \tilta + \aslb^2 \tiltb\right) + \bigo\bigl( \aslb^3 \bigr)\, ,\label{eqn:s0-exp-0}
\end{equation}
\textit{i.e.,}
\begin{align}	
	S_{\anti{f}\anti{X}} &= S_{Xf} \nonumber \\
	  &= \left( S_0 \right)_{Xf} + \imath \left( \aslb \tilta + \aslb^2 \tiltb\right)_{Xf} + \bigo\bigl( \aslb^3 \bigr)
	           \label{eqn:s0-exp-1},
\end{align}
\end{subequations}
where in Eq.~\eqref{eqn:s0-exp-1} we have used CPT conservation, as usual, to rewrite $ S_{\anti{f}\anti{X}} $ as $ S_{Xf} $.

\subsection{\label{ssec:case1}\emph{Case 1}: Where the initial particle decays only by B-violating interactions.}

If the initial particle $X$ and its CP conjugate particle $\anti{X}$ decay only via B-violating interactions, \ie,
\begin{align}
	(S_0)_{\anti{f}\anti{X}} &= (S_0)_{Xf} = \delta_{Xf},\label{eqn:s0_delta-0}
\end{align}
we get, using Eq.~\eqref{eqn:s0_delta-0} in \eqref{eqn:main-eqn2},
\begin{widetext}
\begin{subequations}\label{eqn:cpv-diff}
\begin{align}
	\sumb{f}{B}	
	\abs{\tilt_{Xf}}^2 &= 
	\sumb{f}{B}
	\sum_{j,m} \tilt_{jX}^* S_{jf} \tilt_{mX} S^*_{mf}
	\label{eqn:cpv-diff0}\\
	&=
	\sumb{f}{B} \Biggl(
	\left( \tilt_{fX}^* \tilt_{fX} \right) \Biggr.
	-
	\imath\sum_{m} \tilt_{fX}^* \tilt_{mX} \tilt^*_{mf} 
	+ \imath\sum_{m} \tilt_{mX}^* \tilt_{mf} \tilt_{fX}
	\Biggl. \;+
	\sum_{j,m} \tilt_{jX}^* \tilt_{jf} \tilt_{mX} \tilt^*_{mf} 
	\Biggr),
	\label{eqn:cpv-diff1}
\end{align}	
\end{subequations}
\end{widetext}
where \sumb{f}{B} represents the sum over all final states $f$ with a given baryon number $ B $. In going from Eq.~\eqref{eqn:cpv-diff0} to \eqref{eqn:cpv-diff1} we have expanded $S$ in accordance with Eq.~\eqref{eqn:s-decompose-tilt} and summed over the $\delta_{\alpha\beta}$ as appropriate. We can carry over the first sum in the \emph{R.H.S.}~of Eq.~\eqref{eqn:cpv-diff1} to the other side of the equality, and use CPT as required, to obtain the important difference  in the partial decay widths of the CP conjugate processes violating baryon numbers by $\Delta B = B - B(X)$ and $\Delta \anti{B} = -B - B(\anti{X})$ units respectively.
\begin{equation}
\begin{split}
	&\sumb{\anti{f}}{-B} 
	\abs{\tilt_{\anti{f}\anti{X}}}^2 - \sumb{f}{B} \abs{\tilt_{fX}}^2
	\\
	&\quad
	= 
	\sumb{f}{B}
	\Biggl(
	- \imath\sum_{m} \tilt_{fX}^* \tilt_{mX} \tilt^*_{mf} \\
	&\quad\quad
	+ \imath\sum_{m} \tilt_{fX} \tilt_{mX}^* \tilt_{mf}
	+ \sum_{j,m} \tilt_{jX}^* \tilt_{jf} \tilt_{mX} \tilt^*_{mf}
	\Biggr)
	\label{eqn:cpv-diff2}
\end{split}
\end{equation}

We now expand \tilt~order-by-order in \aslb~according to Eq.~\eqref{eqn:tilt-exp} and evaluate this difference. The results of the calculation to $\bigo(\aslb^2)$ and $\bigo(\aslb^3)$ are enumerated below.
\begin{description}
	\item[To $\bigo(\aslb^2)$:] It is easy to see that each of the three sums in the R.H.S.~of Eq.~\eqref{eqn:cpv-diff2} gives a contribution that is at least to $\bigo(\aslb^3)$. Hence, the $\bigo(\aslb^2)$ contribution to the L.H.S.~is zero. 
Since the tree graph must contain one B-violating vertex, an $\bigo(\aslb^2) $ contribution to the difference in $\abs{T_{fX}}^2$ can only come from the interference of such a tree graph with a loop graph also containing, at most, one B-violating vertex. Thus this result is consistent with the results of the \wn theorem, and shows that graphs to the first order in \aslb~do not contribute to the CP-violating difference.
	
	\item[To $\bigo(\aslb^3)$ and higher:] The $\bigo(\aslb^3)$ contribution to the CP violating difference comes from the first two sums in the R.H.S.,\footnote{The third sum contributes only to $\bigo\bigl( \aslb^4 \bigr)$ and higher.} and is given by
	\begin{equation}
	\begin{split}
		&\sumb{\anti{f}}{-B}		
		\abs{\tilt_{\anti{f}\anti{X}}}^2 - 
		\sumb{f}{B}
		\abs{\tilt_{fX}}^2
		\\
		&\quad=
		\aslb^3 
		\sumb{f}{B}
		\sum_{m}
		2\Im \left((\tilta)_{fX}^* (\tilta)_{mX} (\tilta)^*_{mf}
		     \right)\\
		&\quad\quad
		+ \bigo(\aslb^4)\label{eqn:cpv-leading-sum}
	\end{split}
	\end{equation}
\end{description}

The leading contribution in \aslb~to the CP violating difference is, therefore, to the third order and, as is evident from Eq.~\eqref{eqn:cpv-leading-sum}, comes due to the interference of a tree level graph with its only vertex being B-violating and a loop graph with two B-violating vertices.

\subsection{\label{ssec:case2}\emph{Case 2}: Where the initial particle can decay both through B-conserving and B-violating interactions.}

We now study, in a similar context, the case where the initial particle $X$ may decay via B-conserving as well as B-violating channels to the final states. This translates, in terms of S-matrix elements, to the condition
\begin{equation}\label{eqn:non-trivial-S0}
	(S_0)_{\anti{f}\anti{X}} = (S_0)_{Xf} = \delta_{Xf} + \imath (T_0)_{Xf},
\end{equation}
with $(T_0)_{Xf} \neq 0$, and likewise for $(S_0)_{fX}$. We carry out a similar calculation as in Sec.\ref{ssec:case1}, using Eq.~\eqref{eqn:non-trivial-S0} in \eqref{eqn:main-eqn2}. We expand $\sumb{\anti{f}}{-B}\abs{\tilt_{\anti{f}\anti{X}}}^2 - \sumb{f}{B} \abs{\tilt_{fX}}^2$ order by order in $\aslb$ and read out the $\mathcal{O}(\aslb^2)$ terms in the expansion. Thus, to $\mathcal{O}(\aslb^2)$, we find that 
\begin{widetext}
\begin{equation}
\begin{split}
	\sumb{\anti{f}}{-B} 
	\abs{\tilt_{\anti{f}\anti{X}}}^2 - \sumb{f}{B} \abs{\tilt_{fX}}^2
	&= 
	- \imath\aslb^2
	\sumb{f}{B}
	\left( \tilta \right)_{fX}
	\sum_{m}
	\Bigl(
	(T_0)_{Xm} (\tilta)^*_{fm}
	+ ( \tilta )^*_{mX} ( T_0 )_{mf}
	\Bigr)
	\\
	&\quad+
	\imath\aslb^2
	\sumb{f}{B}
	( \tilta )^*_{fX}
	\sum_{m} 
	\Bigl( (T_0)^*_{mf} (\tilta)_{mX} 
	+
	( T_0 )^*_{Xm}
	( \tilta )_{fm}
	\Bigr).
	\label{eqn:cpv-diff3}
\end{split}
\end{equation}
\end{widetext}
We would now like to compare the CP-violating difference in \emph{case 2} given by Eq.~\eqref{eqn:cpv-diff3} with that found for \emph{case 1} given by Eq.~\eqref{eqn:cpv-leading-sum}. Since, to start with, we have assumed that $(T_0)_{Xm}\neq 0$, and since $T_0$ contains only B conserving interactions, we have $B(X)=B(m)$. But, as B has to be finally violated in the decay of $X$, $B(X)\neq B(f)$. Therefore, we arrive at the conclusion that $B(m)\neq B(f)$ which implies that $(T_0)_{mf}=0=(T_0)^*_{mf}$. Using this result, we find that 
\begin{equation}
\begin{split}
	&\sumb{\anti{f}}{-B}\abs{\tilt_{\anti{f}\anti{X}}}^2
	-
	\sumb{f}{B}	
	\abs{\tilt_{fX}}^2
	\\
	&\quad=
	\aslb^2 
	\sumb{f}{B}
	\sum_{m} 2\Im \left( (\tilta)_{fX}^* (T_0)_{mX}
	(\tilta)^*_{mf}\right)
	\\
	&\quad\quad
	+ \bigo(\aslb^3),
	\label{eqn:cpcv-leading-sum}
\end{split}
\end{equation}
\ie, a non-zero contribution to $\bigo(\aslb^2)$, unlike in Eq.~\eqref{eqn:cpv-leading-sum} where we had only obtained non-zero contributions to $\bigo(\aslb^3)$ and higher. Here we have used the approximate equality of $(\tilta)^*_{ij}$ and $(\tilta)_{ji}$, since their difference is higher order in $\aslb$ and similarly for $(T_0)^*_{ij}$ and $(T_0)_{ji}$. Note the very similar form of  Eq.~\eqref{eqn:cpv-leading-sum} and \eqref{eqn:cpcv-leading-sum}. The important difference, however,  is that a baryon number violating vertex in \emph{case 1} has been replaced by a baryon number conserving vertex in \emph{case 2}, thereby inducing a corresponding change in the transition amplitudes $ \aslb\tilta \to T_0 $ in the respective expressions. Since B/L violating couplings in almost all models are constrained to be small, this replacement allows for the possibility of generating a higher degree of CP violation than would perhaps have been possible with B/L violating decays alone. The example below demonstrates this explicitly.

\section{A toy model in baryogenesis}
\begin{figure*}[htb]
	\includegraphics[width=0.95\textwidth]{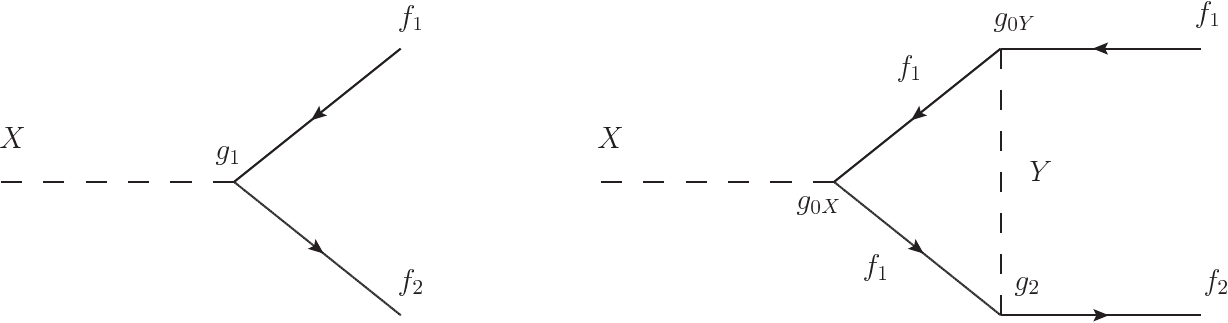} 
	\caption{\label{fig:GUT}Tree level and one-loop diagrams for the decay $X \rightarrow \bar{f_1} + f_2$. Similar diagrams also apply for the decay of the $Y$ boson.}
\end{figure*}

To illustrate the main idea of this paper (i.e., case 2 in Sec IIIB), we consider a toy model for baryogenesis, following an example in Kolb and Turner~\cite{KT}. The model involves two superheavy bosons $X$ and $Y$, whose B-violating out-of-equilibrium decays can generate the necessary B- and CP-violations. For the following discussion we assign baryon numbers of $X$ and $Y$ to be $B(X)=B(Y)=0$. The relevant terms in the Lagrangian are given by
\begin{equation}
\mathcal{L}=g_{0X} X f_1^\dag f_1 + g_{0Y} Y f_1^\dag f_1 + g_1 X f_2^\dag f_1 + g_2 Y f_2^\dag f_1+ {\rm h.c.} .
\end{equation}
Here, $f_i$ ($i = 1, 2$) denote fermions carrying distinct non-zero baryon numbers and equal $U(1)_\text{em}$ charges.
Both bosons have zero $U(1)_\text{em}$ charge.
In the above Lagrangian, $g_{0X}$ and $g_{0Y} $ are B-conserving real couplings, while $g_1$ and $g_2$ are B-violating and complex. Now, consider the B-violating process
\begin{equation}
X \rightarrow \anti{f_1} + f_2,
\end{equation}
where, the leading CP-violating contribution to the decay width comes from the interference of the tree and loop diagrams in figure~\ref{fig:GUT} . These interference terms in the decays of $X$ and $\anti{X}$ are given by
\begin{eqnarray}
\Gamma (X \rightarrow \anti{f_1}+f_2)  &=&  g_1 g_{0X} g_2^* g_{0Y} I_{XY} + {\rm c.c.} \nonumber \\
\Gamma (\anti{X} \rightarrow f_1+\anti{f_2})  &=&  g_1^* g_{0X} g_2 g_{0Y} I_{XY} + {\rm c.c.}, \nonumber \\
\end{eqnarray}
where $I_{XY}$, denoting the loop-factor, can have a non-zero imaginary part when $f_1$ and $f_2$ are lighter than the $X$ boson and can go on-shell inside the loop. The  resulting CP-violation in X-decays will then be
\begin{equation}
\epsilon_X = \frac{4 g_{0X} g_{0Y}\Im(g_1 g_2^*) \Im(I_{XY})}{\Gamma_X},
\end{equation}
which is non-zero in general (here, $\Gamma_X=\Gamma (X \rightarrow \anti{f_1}+f_2) +\Gamma (\anti{X} \rightarrow f_1+\anti{f_2}) $). Similarly, the decays of the Y-boson will lead to a CP-violation as well, which is given by 
\begin{equation}
\epsilon_Y = \frac{4 g_{0X} g_{0Y}\Im(g_2 g_1^*) \Im(I^\prime_{YX})}{\Gamma_Y},
\end{equation}
As long as the $X$ and $Y$ bosons have different masses, the total CP-asymmetry is non-zero, and the resulting B-asymmetry is as follows:
\begin{multline}
\Delta B = (B_2-B_1) \times [4 g_{0X} g_{0Y}\Im(g_1 g_2^*)] \\
\times \left[\frac{\Im(I_{XY})}{\Gamma_X}-\frac{\Im(I^\prime_{YX})}{\Gamma_Y}\right]
\end{multline}
Thus, as expected from our general arguments in the previous section, once a heavy particle has both B-conserving and violating modes of decay, we can generate a B-asymmetry which involves graphs of only first order in B-violation, and therefore, the interference term is only second order in such couplings (in the above example, $\Delta B$ is proportional to $\Im(g_1 g_2^*))$.
In the Appendix, we re-express the standard \wn example in terms of our formulation, where, by considering a boson, X, that does not have any B-conserving decay mode, we verify that the B-asymmetry consequently generated is indeed zero upto second order in B-violation. Therein we also discuss an example from Kolb and Turner \cite{KT}, where additional B-violating decay modes of $X$ help generate an asymmetry at higher orders in B-violation.

\section{A model in low-scale leptogenesis}
\begin{figure*}[htb]
	\includegraphics[width=0.95\textwidth]{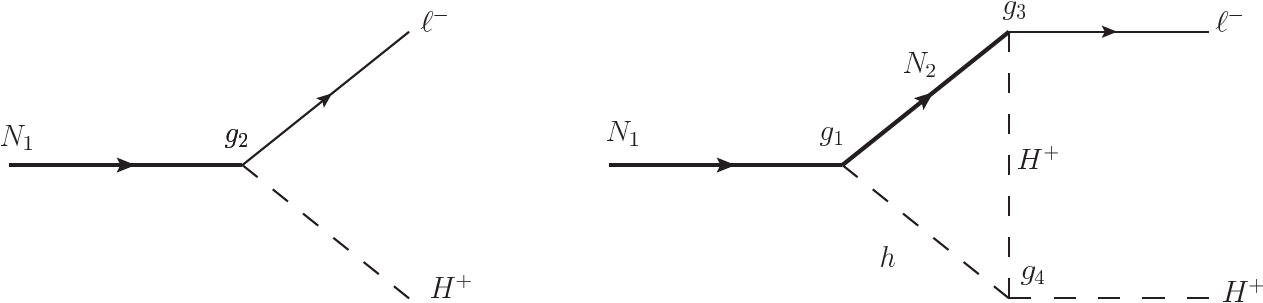} 
	\caption{\label{fig:Ndecay} Tree level and one-loop diagrams for the decay $N_1 \rightarrow \ell^- H^+$.}
\end{figure*}

After considering the above toy model in baryogenesis which demonstrates the primary result of our paper in a very simple example, in this section we give a very brief sketch of a more realistic model in leptogenesis, inspired by the work of Kayser and Segre~\cite{KS}.
In particular, our goal here is to construct an EWSB-scale leptogenesis model utilising both the idea of introducing scalar quartic couplings in the loop graphs as in Ref.~\cite{KS}, as well as having both L-conserving and L-violating decay modes as discussed in case 2 in Sec~\ref{ssec:case2}.

We introduce two right-handed Majorana neutrinos $N_1$ and $N_2$ with masses in the electroweak scale such that $M_{N_1} > M_{N_2}$.
Additionally, we introduce another scalar doublet, $\Phi_2$ (apart from the SM Higgs $\Phi_1$); henceforth, $h$ will represent the SM-like Higgs boson, while $H^+$ will represent the charged Higgs boson from the extended Higgs sector.
This leads to the following possible decay modes for the $N_1$:
\begin{align}
N_1 &\to \ell^-+ H^+ \label{N11} \\
N_1 &\to N_2 + h. \label{N12}
\end{align} 
Consequently, the decay in Eq.~\ref{N11} will arise out of an Yukawa-type interaction and the decay in Eq.~\ref{N12} can arise from a coupling of the form $N_1 N_2 S$ (all of which are SM gauge-singlets) after electroweak symmetry breaking, whereby the singlet $S$ can mix with the neutral components of the doublet scalars.
Due to the Majorana nature of the heavy right-handed neutrinos, depending on the L-number assignment, either the decay in Eq.~\ref{N11}, or its conjugate process ($N_1 \rightarrow \ell^+ H^-$), or both will violate L-number, while the decay in Eq.~\ref{N12} will always be L-conserving (since $L(N_1)=L(N_2))$.

The final ingredient in our model is a quartic coupling between the scalar doublets, of the form $\lambda( \Phi_1^\dag \Phi_1) (\Phi_2^\dag \Phi_2)$, which, after EWSB, will give rise to trilinear scalar couplings.
With this understanding, we consider the diagrams for the process in Eq.~\ref{N11}, as shown in figure~\ref{fig:Ndecay}.
The coupling notations also follow figure~\ref{fig:Ndecay}.
The relevant interference term is given by
\begin{equation}
\Gamma(N_1 \rightarrow \ell^+ H^-) = g_2 g_1 g_3^* g_4 I_{N_1} + c.c.
\end{equation}
Here, the Yukawa couplings $g_2$ and $g_3$ are complex in general, while $g_1$ and $g_4$ are real. The kinematic loop-factor has been denoted by $I_{N_1}$. The resulting CP-violation is then
\begin{equation}
\epsilon_{N_1} = \frac{4 g_1 g_4 \Im(g_2 g_3^*) \Im(I_{N_1})}{\Gamma_{N_1}},
\end{equation} 
where,
\begin{equation}
\begin{split}
	\Gamma_{N_1} &= \Gamma(N_1 \rightarrow \ell^- H^+)+\Gamma(N_1 \rightarrow \ell^+ H^-)\\
	             &  \propto \lvert g_2 \rvert^2.
\end{split}
\end{equation}
Therefore, with the simplifying assumption that $|g_2| \sim |g_3|$, we see that the magnitude of the Yukawa coupling actually cancels out from the CP-violation:
\begin{equation}
\epsilon_{N_1} \simeq \frac{4 g_1 g_4 \delta \Im(I_{N_1})}{(M_{N_1}/8 \pi)},
\label{cp}
\end{equation}
where, the factor $\delta=\sin(\phi_2 - \phi_3)$ comes from the difference of phases of $g_2$ and $g_3$. We have also used $\Gamma_{N_1} \sim |g_2|^2 M_{N_1}/8 \pi$ in writing the above expression. Now, let us estimate the magnitudes of the various terms in Eq.~\ref{cp}:
\begin{enumerate}
\item $g_1$: The $N_1 N_2 S$ coupling is dimensionless, and assumed to be of $\mathcal{O}(1)$. Therefore, $g_1$ is essentially determined by the mixing of the singlet $S$ with the neutral components of the Higgs doublets. For simplicity, we assume that $S$ dominantly mixes with the SM-like lighter Higgs state recently discovered at the LHC. In that case, the measurement of the Higgs properties puts an upper bound on this mixing $\sin \alpha < 10^{-2}$. Hence, we can safely take $g_1 \simeq 10^{-2}$.

\item $g_4$: Since $g_4$ arises after EWSB from a Higgs quartic coupling discussed above, we have $g_4 \sim \lambda v \sin \beta$, with $v=246$ GeV, and $\tan \beta = v_1/v_2$ denotes the ratio of the vacuum expectation values of the neutral CP-even components of $\Phi_1$ and $\Phi_2$, respectively. In a electroweak scale model for leptogenesis, $M_{N_1}$ is also of the order of $v$. And therefore, the factor of $M_{N_1}/8 \pi$ in the denominator will roughly cancel out the factor $v \sin \beta$ in the numerator. 

\item $\Im(I_{N_1})$: This loop factor is found to be
\begin{equation}
	\Im(I_{N_1}) \approx \frac{1}{\pi}\frac{M_{N2}}{M_{N1}^2}\frac{1}{1+\xi^2}\log\left(\frac{1-\xi+\xi^2}{\xi}\right)\,,
\end{equation}
where, $\xi = (M_{H^{\pm}} / M_{N1})^2$. Considering the present constraints on a charged Higgs boson mass, we can safely take $M_{H^+} \simeq 300$ GeV. This leads to a value for the loop factor $\mathcal{O}(10^{-3}-10^{-4})$, for $M_{N_2}<M_{H^+}<M_{N_1}$, and $M_{N_1} \sim 500$ GeV. 

\item $\delta$: This phase factor has a maximum value of 1.
\end{enumerate}
Therefore, for our order of magnitude estimate, we finally obtain
\begin{equation}
\epsilon_{N_1} \sim 10^{-5} \lambda
\end{equation}
For generating a sufficient lepton asymmetry (which is converted to the required baryon asymmetry by the sphaleron processes), one requires $\epsilon_{N_1} = \mathcal {O}(10^{-6})$. Thus we need a quartic coupling of $\lambda = 0.1$, which is a likely value (especially in the light of the recent Higgs mass measurement, whereby the SM Higgs quartic can be estimated to be $\sim 0.13$). 
It is to be noted that the phase factor $\delta$ vanishes if $\phi_2 = \phi_3$. Hence, the couplings of the two RH neutrino mass eigenstates $N_1$ and $N_2$ to the charged lepton $\ell$ should have different phases in order to obtain a nonzero $\epsilon_{N_1}$.

This rather schematic discussion illustrates the feasibility of having models of electroweak scale leptogenesis where the amount of CP-violation is not directly related to the neutrino Yukawa couplings, which, in most low energy (TeV) leptogenesis models, are usually constrained to be small, but rather to the relatively unconstrained quartic Higgs couplings in a two Higgs doublet model. A detailed study of the model is beyond the scope of the present paper, and we leave it to future work.
However, it is important to emphasise the role played by the L-conserving decay channel here --- the absence of such a channel would have entailed one to look for leptogenesis involving graphs with higher order L-violating couplings within the purview of this model, possibly requiring two or more loops and therefore suppressing the generated CP violation significantly.

\section{Remarks and Conclusion}

We have  expanded the interaction amplitude in a perturbation series in the B\slash L-violating coupling \aslb, in order to  show the non-trivial implication of the \wn theorem in the case where B\slash L assignments are naturally and consistently  such that the initial particle may decay by B\slash L-conserving interactions in addition to B\slash L-violating interactions. In particular, it turns out that in such cases, the asymmetry generated due to B\slash L-violating decays may be augmented by B\slash L-conserving interactions in the loop graphs, in a way that deceptively appears contrary to the consequences of the \wn theorem. This re-interpretation of the theorem has significant implications for models of baryogenesis and leptogenesis  by opening up channels which allow for the generation of CP violation that might have been earlier ignored with the intention of subscribing to the theorem's stringent requirements. Additionally, the replacement of a B/L violating coupling by a B/L conserving one, as discussed above, may allow for enhanced generation of CP violation since the former are typically constrained by experiment to be small. As is well known, the generation of ``sufficient'' CP asymmetry remains an issue not just in the Standard Model but in most extensions of it as well. We have illustrated our main result by constructing a toy model in baryogenesis from out-of-equilibrium decays of heavy bosons. 

In addition to setting up new models for B\slash L-genesis employing B\slash L conserving channels as we have shown, it might be an interesting exercise to re-analyse some currently proposed models  of Baryogenesis and Leptogenesis in the light of this interpretation. As an example of this approach, we have considered a recently proposed model of leptogenesis which generates a CP asymmetry only at the two-loop level. By studying a simple variation of this model obtained by slightly altering its particle content in a way  which allows B\slash L conserving decays, we have shown that it is possible to generate sufficient CP asymmetry at the one loop level. 

\acknowledgments
The authors would like to thank Ashoke Sen for his invaluable suggestions in clarifying several key points in the present work and H. S. Mani and M.K. Parida for useful discussions. The authors are also grateful to Boris Kayser for discussions that led them  to the present work.
This work is supported by the World Premier International Research Center Initiative (WPI Initiative), MEXT, Japan, for SM and he would  also like to thank Kaoru Hagiwara for useful discussions and the KEK theory group for warm hospitality in a period during which part of this work was carried out.
This research was supported in part by US Department of Energy contracts DE-FG02-04ER41298 and DE-FG02-13ER41976.
RG also thanks the CERN Theory Division and the University of Wisconsin at Madison phenomenology group for hospitality while the work was in progress.

\appendix
\section{\label{sec:wn_theorem_example}Examples in Baryogenesis to demonstrate the \wn theorem}

Typically, the contribution to baryon asymmetry generated by the particle $X$ with baryon number $B(X) = B_X$ and total decay width $\Gamma_X$, due to its transition to final states $f$ with $B(f) = B_f \neq B_X$, is given by
\begin{equation}
\begin{split}
	\epsilon_X 
	&=
	\sum_f
	\left( B_f - B_X \right)
	\frac{\Gamma(X \to f) - \Gamma(\anti{X} \to \anti{f})}
	     {\Gamma_X}
	\\
	&\propto
	\sum_f
	\left( B_f - B_X \right)
	\sum_m\Im(T^*_{fX}T_{mX}T^*_{mf}).
\end{split}
\end{equation}

\begin{figure*}[htb]
	\centering
	\includegraphics[width=0.95\textwidth]{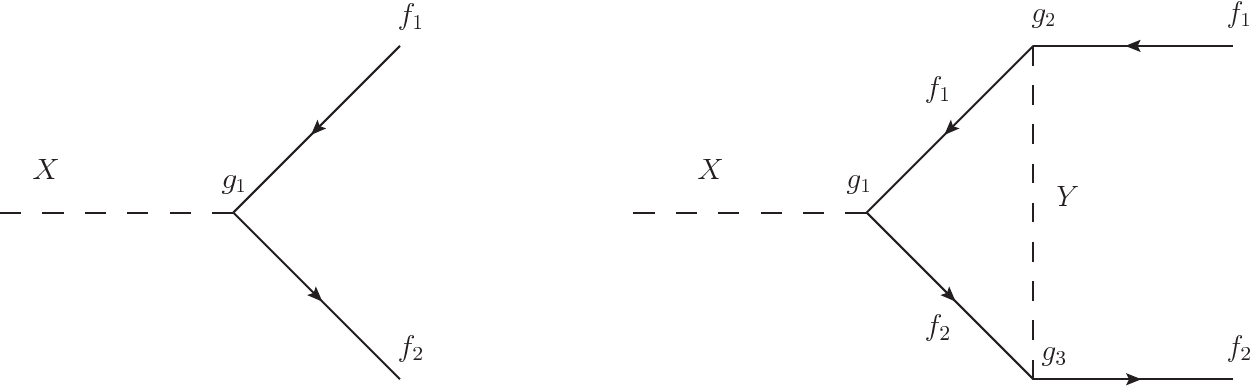}
	\caption{\label{fig:WN_0}Tree and one-loop graphs for the decay $X\to \anti{f}_1 f_2$ due to the Lagrangian $\Lag_a$.}
\end{figure*}

We consider two examples to illustrate the implications of the \wn theorem. First, we consider a model in which a heavy scalar boson $X$ with baryon number $B_X=0$ can decay via a B-violating interaction to a pair of fermions $f_1$ and $f_2$, while another scalar heavy boson $Y$ can decay only via separate B-conserving interactions to both the fermions. The Lagrangian for the model is given below:
\begin{equation}
	\Lag_a = g_1 X f_2^\dag f_1 + g_2 Y f_1^\dag f_1 + g_3 Y f_2^\dag f_2 + h.c.
\end{equation}
The possible tree and one-loop diagrams for the decay process $X\to \anti{f}_1 f_2$ are shown in figure~\ref{fig:WN_0}. Both the tree and one-loop graph have one B-violating vertex each (\textit{vertex with coupling constant $g_1$ in both graphs}).
One can easily calculate the asymmetry generated, $\epsilon_X$, in the decay of $X$ due to the interference of the two graphs and find that
\begin{subequations}
\begin{gather}
	\Gamma(X \to \anti{f}_1 f_2)
	= \abs{g_1}^2 g_2 g_3 
	\left( I_{XY} + I_{XY}^* \right), \text{ and,}
	\\
	\Gamma(\anti{X} \to f_1 \anti{f}_2) = \abs{g_1}^2 g_2 g_3 
	\left( I_{XY} + I_{XY}^* \right),
	\\
	\intertext{which means,}
	\epsilon_{X} \propto
	\Gamma(X \to \anti{f}_1 f_2) -
	\Gamma(\anti{X} \to f_1 \anti{f}_2)
	= 0.\label{eqn:WN0_CPV}
\end{gather}
\end{subequations}
Here, we have represented only the contribution to the decay width arising due to the interference between a one-loop graph and a tree graph by $\Gamma$. The kinematic factor arising out of the integral over the loop-momentum is denoted by $I_{XY}$, which can be complex if the fermions in the loop are kinematically allowed to go on-shell.
As a result of Eq.~\eqref{eqn:WN0_CPV}, the asymmetry generated due to $X$ decays in this model, which is proportional to the CP violation, also becomes zero. This is, clearly, what we expect from the \wn theorem, as the only contributions to the B-violating decay $X\to \anti{f}_1 f_2$ come from processes represented by graphs to the first order in B-violation.

\begin{figure*}[htb]
	\centering
	\includegraphics[width=0.95\textwidth]{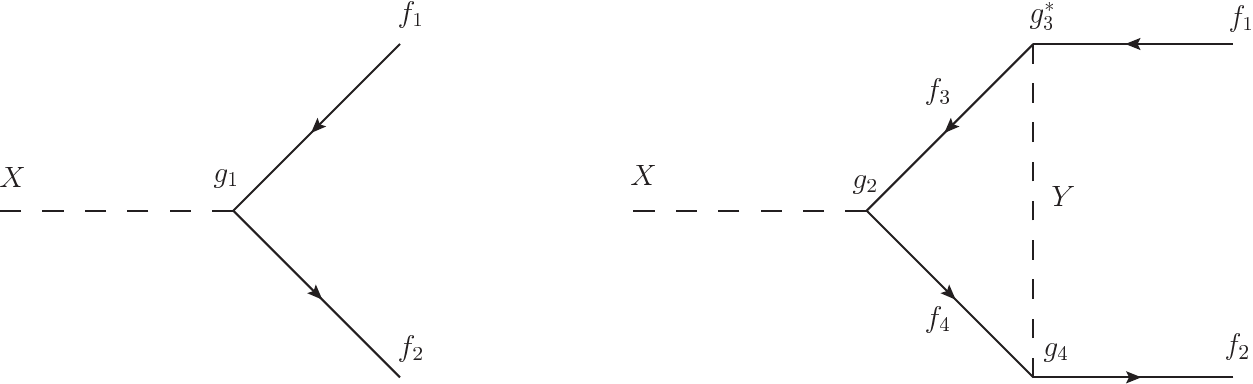}
	\caption{\label{fig:WN_1}Tree and one-loop graphs for the decay $X\to \anti{f}_1 f_2$ due to the Lagrangian $\Lag_b$.}
\end{figure*}

We next consider a model in which both the super-heavy bosons $X$ and $Y$ can decay only via B-violating interactions to fermion pairs. The interaction Lagrangian for this model is given by:
\begin{multline}
	\Lag_b = g_1 X f_2^\dag f_1 + g_2 X f_4^\dag f_3
	\\
	     + g_3 Y f_1^\dag f_3 + g_4 Y f_2^\dag f_4
	     + h.c., 
\end{multline}
where each fermion $f_i$ has a different and unique B-number $B_i$.
The baryon asymmetry generated out of the decays of the super-heavy scalars $X$ and $Y$ in this model has been extensively studied in the literature (see \eg, \cite{KT}). The graphs at the tree and one-loop levels that contribute to the decay $X \to\anti{f}_1 f_2$ are shown in figure~\ref{fig:WN_1}; the loop graph in this case has three B-violating vertices.
It is easy to see that the asymmetry generated in this case is non-zero:
\begin{equation}
	\epsilon_X^{12}
	=
	\frac{4\left( B_2 - B_1 \right)
	\Im \left( I_{XY} \right)
	\Im \left( g_1^* g_2 g_3^* g_4 \right)}
	{\Gamma_X}
	,
\end{equation}
where, as usual, $I_{XY}$ denotes a factor arising out of integration over the loop momentum. One can similarly see that the asymmetry generated due to the decay $X \to \anti{f}_3 f_4$ is given by
\begin{equation}
	\epsilon_X^{34}
	=
	\frac{4\left( B_4 - B_3 \right)
	\Im \left( I_{XY} \right)
	\Im \left( g_1 g_2^* g_3 g_4^* \right)}
	{\Gamma_X}.
\end{equation}
The total asymmetry due to all possible B-violating decays of $X$ is, thus,
\begin{equation}
\begin{split}
	\epsilon_X 
	&= \epsilon_X^{34} + \epsilon_X^{12}\\
	&= 
	\frac{4\bigl( \left(B_4 - B_3\right) -
	       \left(B_2 - B_1\right)
	       \bigr)
	}
	{\Gamma_X}	\\
	& \quad \times
	\Im \left( I_{XY} \right)
	\Im \left( g_1 g_2^* g_3 g_4^* \right)
	\\
	&\neq 0.
\end{split}	
\end{equation}
This is also what is expected from the \wn theorem, since the one-loop contribution to the B-violating decays in this case are of the third order in B-violation.

\bibliography{lepto5}

\begin{thebibliography}{39}%
\makeatletter
\providecommand \@ifxundefined [1]{%
 \@ifx{#1\undefined}
}%
\providecommand \@ifnum [1]{%
 \ifnum #1\expandafter \@firstoftwo
 \else \expandafter \@secondoftwo
 \fi
}%
\providecommand \@ifx [1]{%
 \ifx #1\expandafter \@firstoftwo
 \else \expandafter \@secondoftwo
 \fi
}%
\providecommand \natexlab [1]{#1}%
\providecommand \enquote  [1]{``#1''}%
\providecommand \bibnamefont  [1]{#1}%
\providecommand \bibfnamefont [1]{#1}%
\providecommand \citenamefont [1]{#1}%
\providecommand \href@noop [0]{\@secondoftwo}%
\providecommand \href [0]{\begingroup \@sanitize@url \@href}%
\providecommand \@href[1]{\@@startlink{#1}\@@href}%
\providecommand \@@href[1]{\endgroup#1\@@endlink}%
\providecommand \@sanitize@url [0]{\catcode `\\12\catcode `\$12\catcode
  `\&12\catcode `\#12\catcode `\^12\catcode `\_12\catcode `\%12\relax}%
\providecommand \@@startlink[1]{}%
\providecommand \@@endlink[0]{}%
\providecommand \url  [0]{\begingroup\@sanitize@url \@url }%
\providecommand \@url [1]{\endgroup\@href {#1}{\urlprefix }}%
\providecommand \urlprefix  [0]{URL }%
\providecommand \Eprint [0]{\href }%
\providecommand \doibase [0]{http://dx.doi.org/}%
\providecommand \selectlanguage [0]{\@gobble}%
\providecommand \bibinfo  [0]{\@secondoftwo}%
\providecommand \bibfield  [0]{\@secondoftwo}%
\providecommand \translation [1]{[#1]}%
\providecommand \BibitemOpen [0]{}%
\providecommand \bibitemStop [0]{}%
\providecommand \bibitemNoStop [0]{.\EOS\space}%
\providecommand \EOS [0]{\spacefactor3000\relax}%
\providecommand \BibitemShut  [1]{\csname bibitem#1\endcsname}%
\let\auto@bib@innerbib\@empty
\bibitem [{\citenamefont {Iocco}\ \emph {et~al.}(2009)\citenamefont {Iocco},
  \citenamefont {Mangano}, \citenamefont {Miele}, \citenamefont {Pisanti},\
  and\ \citenamefont {Serpico}}]{Iocco:2008va}%
  \BibitemOpen
  \bibfield  {author} {\bibinfo {author} {\bibfnamefont {F.}~\bibnamefont
  {Iocco}}, \bibinfo {author} {\bibfnamefont {G.}~\bibnamefont {Mangano}},
  \bibinfo {author} {\bibfnamefont {G.}~\bibnamefont {Miele}}, \bibinfo
  {author} {\bibfnamefont {O.}~\bibnamefont {Pisanti}}, \ and\ \bibinfo
  {author} {\bibfnamefont {P.~D.}\ \bibnamefont {Serpico}},\ }\href {\doibase
  10.1016/j.physrep.2009.02.002} {\bibfield  {journal} {\bibinfo  {journal}
  {Phys.Rept.}\ }\textbf {\bibinfo {volume} {472}},\ \bibinfo {pages} {1}
  (\bibinfo {year} {2009})},\ \Eprint {http://arxiv.org/abs/0809.0631}
  {arXiv:0809.0631 [astro-ph]} \BibitemShut {NoStop}%
\bibitem [{\citenamefont {Larson}\ \emph {et~al.}(2011)\citenamefont {Larson},
  \citenamefont {Dunkley}, \citenamefont {Hinshaw}, \citenamefont {Komatsu},
  \citenamefont {Nolta} \emph {et~al.}}]{Larson:2010gs}%
  \BibitemOpen
  \bibfield  {author} {\bibinfo {author} {\bibfnamefont {D.}~\bibnamefont
  {Larson}}, \bibinfo {author} {\bibfnamefont {J.}~\bibnamefont {Dunkley}},
  \bibinfo {author} {\bibfnamefont {G.}~\bibnamefont {Hinshaw}}, \bibinfo
  {author} {\bibfnamefont {E.}~\bibnamefont {Komatsu}}, \bibinfo {author}
  {\bibfnamefont {M.}~\bibnamefont {Nolta}},  \emph {et~al.},\ }\href {\doibase
  10.1088/0067-0049/192/2/16} {\bibfield  {journal} {\bibinfo  {journal}
  {Astrophys.J.Suppl.}\ }\textbf {\bibinfo {volume} {192}},\ \bibinfo {pages}
  {16} (\bibinfo {year} {2011})},\ \Eprint {http://arxiv.org/abs/1001.4635}
  {arXiv:1001.4635 [astro-ph.CO]} \BibitemShut {NoStop}%
\bibitem [{\citenamefont {Steigman}(1976)}]{Steigman:1976ev}%
  \BibitemOpen
  \bibfield  {author} {\bibinfo {author} {\bibfnamefont {G.}~\bibnamefont
  {Steigman}},\ }\href {\doibase 10.1146/annurev.aa.14.090176.002011}
  {\bibfield  {journal} {\bibinfo  {journal} {Ann.Rev.Astron.Astrophys.}\
  }\textbf {\bibinfo {volume} {14}},\ \bibinfo {pages} {339} (\bibinfo {year}
  {1976})}\BibitemShut {NoStop}%
\bibitem [{\citenamefont {Cohen}\ \emph {et~al.}(1998)\citenamefont {Cohen},
  \citenamefont {De~Rujula},\ and\ \citenamefont {Glashow}}]{Cohen:1997ac}%
  \BibitemOpen
  \bibfield  {author} {\bibinfo {author} {\bibfnamefont {A.~G.}\ \bibnamefont
  {Cohen}}, \bibinfo {author} {\bibfnamefont {A.}~\bibnamefont {De~Rujula}}, \
  and\ \bibinfo {author} {\bibfnamefont {S.}~\bibnamefont {Glashow}},\ }\href
  {\doibase 10.1086/305328} {\bibfield  {journal} {\bibinfo  {journal}
  {Astrophys.J.}\ }\textbf {\bibinfo {volume} {495}},\ \bibinfo {pages} {539}
  (\bibinfo {year} {1998})},\ \Eprint {http://arxiv.org/abs/astro-ph/9707087}
  {arXiv:astro-ph/9707087 [astro-ph]} \BibitemShut {NoStop}%
\bibitem [{\citenamefont {Sakharov}(1967)}]{Sak}%
  \BibitemOpen
  \bibfield  {author} {\bibinfo {author} {\bibfnamefont {A.}~\bibnamefont
  {Sakharov}},\ }\href@noop {} {\bibfield  {journal} {\bibinfo  {journal}
  {Pisma Zh.Eksp.Teor.Fiz.}\ }\textbf {\bibinfo {volume} {5}},\ \bibinfo
  {pages} {32} (\bibinfo {year} {1967})}\BibitemShut {NoStop}%
\bibitem [{\citenamefont {Kuzmin}\ \emph {et~al.}(1985)\citenamefont {Kuzmin},
  \citenamefont {Rubakov},\ and\ \citenamefont {Shaposhnikov}}]{Kuzmin:1985mm}%
  \BibitemOpen
  \bibfield  {author} {\bibinfo {author} {\bibfnamefont {V.}~\bibnamefont
  {Kuzmin}}, \bibinfo {author} {\bibfnamefont {V.}~\bibnamefont {Rubakov}}, \
  and\ \bibinfo {author} {\bibfnamefont {M.}~\bibnamefont {Shaposhnikov}},\
  }\href {\doibase 10.1016/0370-2693(85)91028-7} {\bibfield  {journal}
  {\bibinfo  {journal} {Phys.Lett.}\ }\textbf {\bibinfo {volume} {B155}},\
  \bibinfo {pages} {36} (\bibinfo {year} {1985})}\BibitemShut {NoStop}%
\bibitem [{Note1()}]{Note1}%
  \BibitemOpen
  \bibinfo {note} {C, P and T will denote the charge-conjugation, parity and
  time-reversal transformations, respectively, henceforth in our
  work.}\BibitemShut {Stop}%
\bibitem [{\citenamefont {Kobayashi}\ and\ \citenamefont
  {Maskawa}(1973)}]{Kobayashi:1973fv}%
  \BibitemOpen
  \bibfield  {author} {\bibinfo {author} {\bibfnamefont {M.}~\bibnamefont
  {Kobayashi}}\ and\ \bibinfo {author} {\bibfnamefont {T.}~\bibnamefont
  {Maskawa}},\ }\href {\doibase 10.1143/PTP.49.652} {\bibfield  {journal}
  {\bibinfo  {journal} {Prog.Theor.Phys.}\ }\textbf {\bibinfo {volume} {49}},\
  \bibinfo {pages} {652} (\bibinfo {year} {1973})}\BibitemShut {NoStop}%
\bibitem [{\citenamefont {Huet}\ and\ \citenamefont
  {Sather}(1995)}]{Huet:1994jb}%
  \BibitemOpen
  \bibfield  {author} {\bibinfo {author} {\bibfnamefont {P.}~\bibnamefont
  {Huet}}\ and\ \bibinfo {author} {\bibfnamefont {E.}~\bibnamefont {Sather}},\
  }\href {\doibase 10.1103/PhysRevD.51.379} {\bibfield  {journal} {\bibinfo
  {journal} {Phys.Rev.}\ }\textbf {\bibinfo {volume} {D51}},\ \bibinfo {pages}
  {379} (\bibinfo {year} {1995})},\ \Eprint
  {http://arxiv.org/abs/hep-ph/9404302} {arXiv:hep-ph/9404302 [hep-ph]}
  \BibitemShut {NoStop}%
\bibitem [{\citenamefont {Rubakov}\ and\ \citenamefont
  {Shaposhnikov}(1996)}]{Rubakov:1996vz}%
  \BibitemOpen
  \bibfield  {author} {\bibinfo {author} {\bibfnamefont {V.}~\bibnamefont
  {Rubakov}}\ and\ \bibinfo {author} {\bibfnamefont {M.}~\bibnamefont
  {Shaposhnikov}},\ }\href {\doibase 10.1070/PU1996v039n05ABEH000145}
  {\bibfield  {journal} {\bibinfo  {journal} {Usp.Fiz.Nauk}\ }\textbf {\bibinfo
  {volume} {166}},\ \bibinfo {pages} {493} (\bibinfo {year} {1996})},\ \Eprint
  {http://arxiv.org/abs/hep-ph/9603208} {arXiv:hep-ph/9603208 [hep-ph]}
  \BibitemShut {NoStop}%
\bibitem [{\citenamefont {Kajantie}\ \emph {et~al.}(1996)\citenamefont
  {Kajantie}, \citenamefont {Laine}, \citenamefont {Rummukainen},\ and\
  \citenamefont {Shaposhnikov}}]{Kajantie:1995kf}%
  \BibitemOpen
  \bibfield  {author} {\bibinfo {author} {\bibfnamefont {K.}~\bibnamefont
  {Kajantie}}, \bibinfo {author} {\bibfnamefont {M.}~\bibnamefont {Laine}},
  \bibinfo {author} {\bibfnamefont {K.}~\bibnamefont {Rummukainen}}, \ and\
  \bibinfo {author} {\bibfnamefont {M.~E.}\ \bibnamefont {Shaposhnikov}},\
  }\href {\doibase 10.1016/0550-3213(96)00052-1} {\bibfield  {journal}
  {\bibinfo  {journal} {Nucl.Phys.}\ }\textbf {\bibinfo {volume} {B466}},\
  \bibinfo {pages} {189} (\bibinfo {year} {1996})},\ \Eprint
  {http://arxiv.org/abs/hep-lat/9510020} {arXiv:hep-lat/9510020 [hep-lat]}
  \BibitemShut {NoStop}%
\bibitem [{\citenamefont {Cline}(2006)}]{Cline:2006ts}%
  \BibitemOpen
  \bibfield  {author} {\bibinfo {author} {\bibfnamefont {J.~M.}\ \bibnamefont
  {Cline}},\ }\href@noop {} {\  (\bibinfo {year} {2006})},\ \Eprint
  {http://arxiv.org/abs/hep-ph/0609145} {arXiv:hep-ph/0609145 [hep-ph]}
  \BibitemShut {NoStop}%
\bibitem [{\citenamefont {Riotto}\ and\ \citenamefont
  {Trodden}(1999)}]{Riotto:1999yt}%
  \BibitemOpen
  \bibfield  {author} {\bibinfo {author} {\bibfnamefont {A.}~\bibnamefont
  {Riotto}}\ and\ \bibinfo {author} {\bibfnamefont {M.}~\bibnamefont
  {Trodden}},\ }\href {\doibase 10.1146/annurev.nucl.49.1.35} {\bibfield
  {journal} {\bibinfo  {journal} {Ann.Rev.Nucl.Part.Sci.}\ }\textbf {\bibinfo
  {volume} {49}},\ \bibinfo {pages} {35} (\bibinfo {year} {1999})},\ \Eprint
  {http://arxiv.org/abs/hep-ph/9901362} {arXiv:hep-ph/9901362 [hep-ph]}
  \BibitemShut {NoStop}%
\bibitem [{\citenamefont {Riotto}(1998)}]{Riotto:1998bt}%
  \BibitemOpen
  \bibfield  {author} {\bibinfo {author} {\bibfnamefont {A.}~\bibnamefont
  {Riotto}},\ }\href@noop {} {\ ,\ \bibinfo {pages} {326} (\bibinfo {year}
  {1998})},\ \Eprint {http://arxiv.org/abs/hep-ph/9807454}
  {arXiv:hep-ph/9807454 [hep-ph]} \BibitemShut {NoStop}%
\bibitem [{\citenamefont {Ignatiev}\ \emph {et~al.}(1978)\citenamefont
  {Ignatiev}, \citenamefont {Krasnikov}, \citenamefont {Kuzmin},\ and\
  \citenamefont {Tavkhelidze}}]{Ignatiev:1978uf}%
  \BibitemOpen
  \bibfield  {author} {\bibinfo {author} {\bibfnamefont {A.~Y.}\ \bibnamefont
  {Ignatiev}}, \bibinfo {author} {\bibfnamefont {N.}~\bibnamefont {Krasnikov}},
  \bibinfo {author} {\bibfnamefont {V.}~\bibnamefont {Kuzmin}}, \ and\ \bibinfo
  {author} {\bibfnamefont {A.}~\bibnamefont {Tavkhelidze}},\ }\href {\doibase
  10.1016/0370-2693(78)90900-0} {\bibfield  {journal} {\bibinfo  {journal}
  {Phys.Lett.}\ }\textbf {\bibinfo {volume} {B76}},\ \bibinfo {pages} {436}
  (\bibinfo {year} {1978})}\BibitemShut {NoStop}%
\bibitem [{\citenamefont {Yoshimura}(1978)}]{Yoshimura:1978ex}%
  \BibitemOpen
  \bibfield  {author} {\bibinfo {author} {\bibfnamefont {M.}~\bibnamefont
  {Yoshimura}},\ }\href {\doibase 10.1103/PhysRevLett.41.281} {\bibfield
  {journal} {\bibinfo  {journal} {Phys.Rev.Lett.}\ }\textbf {\bibinfo {volume}
  {41}},\ \bibinfo {pages} {281} (\bibinfo {year} {1978})}\BibitemShut
  {NoStop}%
\bibitem [{\citenamefont {Toussaint}\ \emph {et~al.}(1979)\citenamefont
  {Toussaint}, \citenamefont {Treiman}, \citenamefont {Wilczek},\ and\
  \citenamefont {Zee}}]{Toussaint:1978br}%
  \BibitemOpen
  \bibfield  {author} {\bibinfo {author} {\bibfnamefont {D.}~\bibnamefont
  {Toussaint}}, \bibinfo {author} {\bibfnamefont {S.}~\bibnamefont {Treiman}},
  \bibinfo {author} {\bibfnamefont {F.}~\bibnamefont {Wilczek}}, \ and\
  \bibinfo {author} {\bibfnamefont {A.}~\bibnamefont {Zee}},\ }\href {\doibase
  10.1103/PhysRevD.19.1036} {\bibfield  {journal} {\bibinfo  {journal}
  {Phys.Rev.}\ }\textbf {\bibinfo {volume} {D19}},\ \bibinfo {pages} {1036}
  (\bibinfo {year} {1979})}\BibitemShut {NoStop}%
\bibitem [{\citenamefont {Dimopoulos}\ and\ \citenamefont
  {Susskind}(1978)}]{Dimopoulos:1978kv}%
  \BibitemOpen
  \bibfield  {author} {\bibinfo {author} {\bibfnamefont {S.}~\bibnamefont
  {Dimopoulos}}\ and\ \bibinfo {author} {\bibfnamefont {L.}~\bibnamefont
  {Susskind}},\ }\href {\doibase 10.1103/PhysRevD.18.4500} {\bibfield
  {journal} {\bibinfo  {journal} {Phys.Rev.}\ }\textbf {\bibinfo {volume}
  {D18}},\ \bibinfo {pages} {4500} (\bibinfo {year} {1978})}\BibitemShut
  {NoStop}%
\bibitem [{\citenamefont {Ellis}\ \emph {et~al.}(1979)\citenamefont {Ellis},
  \citenamefont {Gaillard},\ and\ \citenamefont {Nanopoulos}}]{Ellis:1978xg}%
  \BibitemOpen
  \bibfield  {author} {\bibinfo {author} {\bibfnamefont {J.~R.}\ \bibnamefont
  {Ellis}}, \bibinfo {author} {\bibfnamefont {M.~K.}\ \bibnamefont {Gaillard}},
  \ and\ \bibinfo {author} {\bibfnamefont {D.~V.}\ \bibnamefont {Nanopoulos}},\
  }\href {\doibase 10.1016/0370-2693(79)91190-0} {\bibfield  {journal}
  {\bibinfo  {journal} {Phys.Lett.}\ }\textbf {\bibinfo {volume} {B80}},\
  \bibinfo {pages} {360} (\bibinfo {year} {1979})}\BibitemShut {NoStop}%
\bibitem [{\citenamefont {Weinberg}(1979)}]{Weinberg:1979bt}%
  \BibitemOpen
  \bibfield  {author} {\bibinfo {author} {\bibfnamefont {S.}~\bibnamefont
  {Weinberg}},\ }\href {\doibase 10.1103/PhysRevLett.42.850} {\bibfield
  {journal} {\bibinfo  {journal} {Phys.Rev.Lett.}\ }\textbf {\bibinfo {volume}
  {42}},\ \bibinfo {pages} {850} (\bibinfo {year} {1979})}\BibitemShut
  {NoStop}%
\bibitem [{\citenamefont {Yoshimura}(1979)}]{Yoshimura:1979gy}%
  \BibitemOpen
  \bibfield  {author} {\bibinfo {author} {\bibfnamefont {M.}~\bibnamefont
  {Yoshimura}},\ }\href {\doibase 10.1016/0370-2693(79)90471-4} {\bibfield
  {journal} {\bibinfo  {journal} {Phys.Lett.}\ }\textbf {\bibinfo {volume}
  {B88}},\ \bibinfo {pages} {294} (\bibinfo {year} {1979})}\BibitemShut
  {NoStop}%
\bibitem [{\citenamefont {Barr}\ \emph {et~al.}(1979)\citenamefont {Barr},
  \citenamefont {Segre},\ and\ \citenamefont {Weldon}}]{Barr:1979ye}%
  \BibitemOpen
  \bibfield  {author} {\bibinfo {author} {\bibfnamefont {S.~M.}\ \bibnamefont
  {Barr}}, \bibinfo {author} {\bibfnamefont {G.}~\bibnamefont {Segre}}, \ and\
  \bibinfo {author} {\bibfnamefont {H.~A.}\ \bibnamefont {Weldon}},\ }\href
  {\doibase 10.1103/PhysRevD.20.2494} {\bibfield  {journal} {\bibinfo
  {journal} {Phys.Rev.}\ }\textbf {\bibinfo {volume} {D20}},\ \bibinfo {pages}
  {2494} (\bibinfo {year} {1979})}\BibitemShut {NoStop}%
\bibitem [{\citenamefont {Nanopoulos}\ and\ \citenamefont
  {Weinberg}(1979)}]{WN}%
  \BibitemOpen
  \bibfield  {author} {\bibinfo {author} {\bibfnamefont {D.~V.}\ \bibnamefont
  {Nanopoulos}}\ and\ \bibinfo {author} {\bibfnamefont {S.}~\bibnamefont
  {Weinberg}},\ }\href {\doibase 10.1103/PhysRevD.20.2484} {\bibfield
  {journal} {\bibinfo  {journal} {Phys.Rev.}\ }\textbf {\bibinfo {volume}
  {D20}},\ \bibinfo {pages} {2484} (\bibinfo {year} {1979})}\BibitemShut
  {NoStop}%
\bibitem [{\citenamefont {Yildiz}\ and\ \citenamefont
  {Cox}(1980)}]{Yildiz:1979gx}%
  \BibitemOpen
  \bibfield  {author} {\bibinfo {author} {\bibfnamefont {A.}~\bibnamefont
  {Yildiz}}\ and\ \bibinfo {author} {\bibfnamefont {P.~H.}\ \bibnamefont
  {Cox}},\ }\href {\doibase 10.1103/PhysRevD.21.906} {\bibfield  {journal}
  {\bibinfo  {journal} {Phys.Rev.}\ }\textbf {\bibinfo {volume} {D21}},\
  \bibinfo {pages} {906} (\bibinfo {year} {1980})}\BibitemShut {NoStop}%
\bibitem [{\citenamefont {Affleck}\ and\ \citenamefont
  {Dine}(1985)}]{Affleck:1984fy}%
  \BibitemOpen
  \bibfield  {author} {\bibinfo {author} {\bibfnamefont {I.}~\bibnamefont
  {Affleck}}\ and\ \bibinfo {author} {\bibfnamefont {M.}~\bibnamefont {Dine}},\
  }\href {\doibase 10.1016/0550-3213(85)90021-5} {\bibfield  {journal}
  {\bibinfo  {journal} {Nucl.Phys.}\ }\textbf {\bibinfo {volume} {B249}},\
  \bibinfo {pages} {361} (\bibinfo {year} {1985})}\BibitemShut {NoStop}%
\bibitem [{\citenamefont {Cohen}\ and\ \citenamefont
  {Kaplan}(1987)}]{Cohen:1987vi}%
  \BibitemOpen
  \bibfield  {author} {\bibinfo {author} {\bibfnamefont {A.~G.}\ \bibnamefont
  {Cohen}}\ and\ \bibinfo {author} {\bibfnamefont {D.~B.}\ \bibnamefont
  {Kaplan}},\ }\href {\doibase 10.1016/0370-2693(87)91369-4} {\bibfield
  {journal} {\bibinfo  {journal} {Phys.Lett.}\ }\textbf {\bibinfo {volume}
  {B199}},\ \bibinfo {pages} {251} (\bibinfo {year} {1987})}\BibitemShut
  {NoStop}%
\bibitem [{\citenamefont {Cohen}\ and\ \citenamefont
  {Kaplan}(1988)}]{Cohen:1988kt}%
  \BibitemOpen
  \bibfield  {author} {\bibinfo {author} {\bibfnamefont {A.~G.}\ \bibnamefont
  {Cohen}}\ and\ \bibinfo {author} {\bibfnamefont {D.~B.}\ \bibnamefont
  {Kaplan}},\ }\href {\doibase 10.1016/0550-3213(88)90134-4} {\bibfield
  {journal} {\bibinfo  {journal} {Nucl.Phys.}\ }\textbf {\bibinfo {volume}
  {B308}},\ \bibinfo {pages} {913} (\bibinfo {year} {1988})}\BibitemShut
  {NoStop}%
\bibitem [{\citenamefont {Fukugita}\ and\ \citenamefont {Yanagida}(1986)}]{FY}%
  \BibitemOpen
  \bibfield  {author} {\bibinfo {author} {\bibfnamefont {M.}~\bibnamefont
  {Fukugita}}\ and\ \bibinfo {author} {\bibfnamefont {T.}~\bibnamefont
  {Yanagida}},\ }\href {\doibase 10.1016/0370-2693(86)91126-3} {\bibfield
  {journal} {\bibinfo  {journal} {Phys.Lett.}\ }\textbf {\bibinfo {volume}
  {B174}},\ \bibinfo {pages} {45} (\bibinfo {year} {1986})}\BibitemShut
  {NoStop}%
\bibitem [{\citenamefont {Fong}\ \emph {et~al.}(2012)\citenamefont {Fong},
  \citenamefont {Nardi},\ and\ \citenamefont {Riotto}}]{Fong:2013wr}%
  \BibitemOpen
  \bibfield  {author} {\bibinfo {author} {\bibfnamefont {C.~S.}\ \bibnamefont
  {Fong}}, \bibinfo {author} {\bibfnamefont {E.}~\bibnamefont {Nardi}}, \ and\
  \bibinfo {author} {\bibfnamefont {A.}~\bibnamefont {Riotto}},\ }\href
  {\doibase 10.1155/2012/158303} {\bibfield  {journal} {\bibinfo  {journal}
  {Adv.High Energy Phys.}\ }\textbf {\bibinfo {volume} {2012}},\ \bibinfo
  {pages} {158303} (\bibinfo {year} {2012})},\ \Eprint
  {http://arxiv.org/abs/1301.3062} {arXiv:1301.3062 [hep-ph]} \BibitemShut
  {NoStop}%
\bibitem [{\citenamefont {Nardi}(2013)}]{Nardi:2013mha}%
  \BibitemOpen
  \bibfield  {author} {\bibinfo {author} {\bibfnamefont {E.}~\bibnamefont
  {Nardi}},\ }\href {\doibase 10.1142/9789814436830_0052} {\ ,\ \bibinfo
  {pages} {238} (\bibinfo {year} {2013})}\BibitemShut {NoStop}%
\bibitem [{\citenamefont {Pilaftsis}(2013)}]{Pilaftsis:2013nqa}%
  \BibitemOpen
  \bibfield  {author} {\bibinfo {author} {\bibfnamefont {A.}~\bibnamefont
  {Pilaftsis}},\ }\href {\doibase 10.1088/1742-6596/447/1/012007} {\bibfield
  {journal} {\bibinfo  {journal} {J.Phys.Conf.Ser.}\ }\textbf {\bibinfo
  {volume} {447}},\ \bibinfo {pages} {012007} (\bibinfo {year}
  {2013})}\BibitemShut {NoStop}%
\bibitem [{\citenamefont {Pilaftsis}(2009)}]{Pilaftsis:2009pk}%
  \BibitemOpen
  \bibfield  {author} {\bibinfo {author} {\bibfnamefont {A.}~\bibnamefont
  {Pilaftsis}},\ }\href {\doibase 10.1088/1742-6596/171/1/012017} {\bibfield
  {journal} {\bibinfo  {journal} {J.Phys.Conf.Ser.}\ }\textbf {\bibinfo
  {volume} {171}},\ \bibinfo {pages} {012017} (\bibinfo {year} {2009})},\
  \Eprint {http://arxiv.org/abs/0904.1182} {arXiv:0904.1182 [hep-ph]}
  \BibitemShut {NoStop}%
\bibitem [{\citenamefont {Di~Bari}(2012)}]{DiBari:2012fz}%
  \BibitemOpen
  \bibfield  {author} {\bibinfo {author} {\bibfnamefont {P.}~\bibnamefont
  {Di~Bari}},\ }\href {\doibase 10.1080/00107514.2012.701096} {\bibfield
  {journal} {\bibinfo  {journal} {Contemp.Phys.}\ }\textbf {\bibinfo {volume}
  {53}},\ \bibinfo {pages} {ISSUE4} (\bibinfo {year} {2012})},\ \Eprint
  {http://arxiv.org/abs/1206.3168} {arXiv:1206.3168 [hep-ph]} \BibitemShut
  {NoStop}%
\bibitem [{\citenamefont {Davidson}\ \emph {et~al.}(2008)\citenamefont
  {Davidson}, \citenamefont {Nardi},\ and\ \citenamefont {Nir}}]{DNN}%
  \BibitemOpen
  \bibfield  {author} {\bibinfo {author} {\bibfnamefont {S.}~\bibnamefont
  {Davidson}}, \bibinfo {author} {\bibfnamefont {E.}~\bibnamefont {Nardi}}, \
  and\ \bibinfo {author} {\bibfnamefont {Y.}~\bibnamefont {Nir}},\ }\href
  {\doibase 10.1016/j.physrep.2008.06.002} {\bibfield  {journal} {\bibinfo
  {journal} {Phys.Rept.}\ }\textbf {\bibinfo {volume} {466}},\ \bibinfo {pages}
  {105} (\bibinfo {year} {2008})},\ \Eprint {http://arxiv.org/abs/0802.2962}
  {arXiv:0802.2962 [hep-ph]} \BibitemShut {NoStop}%
\bibitem [{\citenamefont {Buchmuller}\ \emph {et~al.}(2005)\citenamefont
  {Buchmuller}, \citenamefont {Peccei},\ and\ \citenamefont {Yanagida}}]{BPY}%
  \BibitemOpen
  \bibfield  {author} {\bibinfo {author} {\bibfnamefont {W.}~\bibnamefont
  {Buchmuller}}, \bibinfo {author} {\bibfnamefont {R.}~\bibnamefont {Peccei}},
  \ and\ \bibinfo {author} {\bibfnamefont {T.}~\bibnamefont {Yanagida}},\
  }\href {\doibase 10.1146/annurev.nucl.55.090704.151558} {\bibfield  {journal}
  {\bibinfo  {journal} {Ann.Rev.Nucl.Part.Sci.}\ }\textbf {\bibinfo {volume}
  {55}},\ \bibinfo {pages} {311} (\bibinfo {year} {2005})},\ \Eprint
  {http://arxiv.org/abs/hep-ph/0502169} {arXiv:hep-ph/0502169 [hep-ph]}
  \BibitemShut {NoStop}%
\bibitem [{\citenamefont {Kolb}\ and\ \citenamefont {Wolfram}(1980)}]{KW}%
  \BibitemOpen
  \bibfield  {author} {\bibinfo {author} {\bibfnamefont {E.~W.}\ \bibnamefont
  {Kolb}}\ and\ \bibinfo {author} {\bibfnamefont {S.}~\bibnamefont {Wolfram}},\
  }\href {\doibase 10.1016/0550-3213(80)90167-4, 10.1016/0550-3213(80)90167-4}
  {\bibfield  {journal} {\bibinfo  {journal} {Nucl.Phys.}\ }\textbf {\bibinfo
  {volume} {B172}},\ \bibinfo {pages} {224} (\bibinfo {year}
  {1980})}\BibitemShut {NoStop}%
\bibitem [{Note2()}]{Note2}%
  \BibitemOpen
  \bibinfo {note} {The third sum contributes only to $\protect \ensuremath
  {\protect \mathcal {O}}\mathopen {\setbox \z@ \hbox {\frozen@everymath
  \@emptytoks \mathsurround \z@ $\nulldelimiterspace \z@ \left (\vcenter to\@ne
  \big@size {}\right .$}\box \z@ } \protect \ensuremath {\alpha _\protect
  \ensuremath {{\mathchoice {{\setbox \z@ \hbox {$\mathsurround \z@
  \displaystyle B$}\setbox \tw@ \hbox {$\mathsurround \z@ \displaystyle
  /$}\dimen 4\wd \z@ \dimen@ \ht \tw@ \advance \dimen@ -\dp \tw@ \advance
  \dimen@ -\ht \z@ \advance \dimen@ \dp \z@ \divide \dimen@ \tw@ \advance
  \dimen@ -0\ht \tw@ \advance \dimen@ -0\dp \tw@ \dimen@ii 0\wd \z@ \raise
  -\dimen@ \hbox to\dimen 4{\hss \kern \dimen@ii \box \tw@ \kern -\dimen@ii
  \hss }\hbox to\z@ {\hss \hbox to\dimen 4{\hss \box \z@ \hss }}}}{{\setbox \z@
  \hbox {$\mathsurround \z@ \textstyle B$}\setbox \tw@ \hbox {$\mathsurround
  \z@ \textstyle /$}\dimen 4\wd \z@ \dimen@ \ht \tw@ \advance \dimen@ -\dp \tw@
  \advance \dimen@ -\ht \z@ \advance \dimen@ \dp \z@ \divide \dimen@ \tw@
  \advance \dimen@ -0\ht \tw@ \advance \dimen@ -0\dp \tw@ \dimen@ii 0\wd \z@
  \raise -\dimen@ \hbox to\dimen 4{\hss \kern \dimen@ii \box \tw@ \kern
  -\dimen@ii \hss }\hbox to\z@ {\hss \hbox to\dimen 4{\hss \box \z@ \hss
  }}}}{{\setbox \z@ \hbox {$\mathsurround \z@ \scriptstyle B$}\setbox \tw@
  \hbox {$\mathsurround \z@ \scriptstyle /$}\dimen 4\wd \z@ \dimen@ \ht \tw@
  \advance \dimen@ -\dp \tw@ \advance \dimen@ -\ht \z@ \advance \dimen@ \dp \z@
  \divide \dimen@ \tw@ \advance \dimen@ -0\ht \tw@ \advance \dimen@ -0\dp \tw@
  \dimen@ii 0\wd \z@ \raise -\dimen@ \hbox to\dimen 4{\hss \kern \dimen@ii \box
  \tw@ \kern -\dimen@ii \hss }\hbox to\z@ {\hss \hbox to\dimen 4{\hss \box \z@
  \hss }}}}{{\setbox \z@ \hbox {$\mathsurround \z@ \scriptscriptstyle
  B$}\setbox \tw@ \hbox {$\mathsurround \z@ \scriptscriptstyle /$}\dimen 4\wd
  \z@ \dimen@ \ht \tw@ \advance \dimen@ -\dp \tw@ \advance \dimen@ -\ht \z@
  \advance \dimen@ \dp \z@ \divide \dimen@ \tw@ \advance \dimen@ -0\ht \tw@
  \advance \dimen@ -0\dp \tw@ \dimen@ii 0\wd \z@ \raise -\dimen@ \hbox to\dimen
  4{\hss \kern \dimen@ii \box \tw@ \kern -\dimen@ii \hss }\hbox to\z@ {\hss
  \hbox to\dimen 4{\hss \box \z@ \hss }}}}}}}^4 \mathclose {\setbox \z@ \hbox
  {\frozen@everymath \@emptytoks \mathsurround \z@ $\nulldelimiterspace \z@
  \left )\vcenter to\@ne \big@size {}\right .$}\box \z@ }$ and
  higher.}\BibitemShut {Stop}%
\bibitem [{\citenamefont {Kolb}\ and\ \citenamefont {Turner}(1993)}]{KT}%
  \BibitemOpen
  \bibfield  {author} {\bibinfo {author} {\bibfnamefont {E.~W.}\ \bibnamefont
  {Kolb}}\ and\ \bibinfo {author} {\bibfnamefont {M.~S.}\ \bibnamefont
  {Turner}},\ }\enquote {\bibinfo {title} {The early universe},}\ \ (\bibinfo
  {publisher} {WestView Press},\ \bibinfo {year} {1993})\ Chap.~\bibinfo
  {chapter} {6}\BibitemShut {NoStop}%
\bibitem [{\citenamefont {Kayser}\ and\ \citenamefont {Segre}(2010)}]{KS}%
  \BibitemOpen
  \bibfield  {author} {\bibinfo {author} {\bibfnamefont {B.}~\bibnamefont
  {Kayser}}\ and\ \bibinfo {author} {\bibfnamefont {G.}~\bibnamefont {Segre}},\
  }\href@noop {} {\  (\bibinfo {year} {2010})},\ \Eprint
  {http://arxiv.org/abs/1011.6362v1} {arXiv:1011.6362v1 [hep-ph]} \BibitemShut
  {NoStop}%
\end{thebibliography}%

\end{document}